\documentclass[aps,prb,preprint,showpacs,superscriptaddress]{revtex4}
\usepackage{graphicx}  
\usepackage{dcolumn}   
\usepackage{bm}        
\usepackage{amssymb}   

\begin{document}

\title{Theoretical treatment of anharmonicity of vibrational modes of single-walled carbon nanotubes}

\author{Hengjia Wang}
\affiliation{Department of Physics and Astronomy, Clemson University,
  SC, USA 29634} 
\author{Doyl Dickel}
\affiliation{Center for Advanced Vehicular Systems, Mississippi State University, Starkville, MS, USA 39759}
\author{Murray S. Daw}
\affiliation{Department of Physics and Astronomy, Clemson University,
  SC, USA 29634} 

\begin{abstract}
  
We report a computational study, using the ``moments method'' [Y. Gao and M. Daw, \textbf{Modelling Simul. Mater. Sci. Eng. 23} 045002 (2015)], of the anharmonicity of the vibrational modes of single-walled carbon nanotubes. We find that modes with displacements largely within the wall are more anharmonic than modes with dominantly radial character, except for a set of modes that are related to the radial breathing mode which are the most anharmonic of all. We also find that periodicity of the calculation along the tube length does not strongly affect the anharmonicity of the modes, but that the tubes with larger diameter show more anharmonicity. Comparison is made with available experiments and other calculations.

\end{abstract}

\maketitle

\section{Introduction}

Vibrational modes of single-wall carbon nanotubes (SWNTs) are accessible experimentally via Raman spectroscopy.\cite{Hiura1993,Dresselhaus2005,Thomsen2007} A typical Raman spectrum of a SWNT reveals two prominent peaks associated with intrinsic (not defect-related) modes: the Radial Breathing Mode (RBM) and the so-called ``G-band'', a group of high-frequency modes that have an analog in graphene. Extensive theoretical investigations of the vibrational modes of SWNTs have been carried out, most commonly using the harmonic approximation.\cite{Eklund1995,Aydin2010} 

By examining the temperature-dependence of the Raman spectra, it is possible to explore the anharmonicity of the RBM and G-band in particular. Raman spectra of the RBM and G-band in SWNT bundles have revealed that their frequency drops with temperature between $300K$ and $800K$.\cite{HDLi2000,Raravikar2002}. Molecular dynamics calculations on individual SWNTs agree with these observations.\cite{Raravikar2002}

In this work, we apply the ``moments method''\cite{momentstheory, jazzpaper, Jazznote} to investigate the temperature-dependence of the frequency of all of the modes of a SWNT. We also study how the chirality, diameter and length of the tube affect the anharmonicity. We find that generally all of the vibrational modes shift to lower frequency with increasing temperature. Within that generality, we find that, with respect to anharmonicity, there are three basic groups of modes. Vibrational modes with largely longitudinal character (that is, displacements parallel to the length of the tube) as well as modes with largely azimuthal displacements (transverse to the length but tangential to the tube) are more anharmonic than the large majority of radial modes. The exception are the azimuthally symmetric radial modes, which are the most anharmonic of all modes. The RBM itself is the most anharmonic of all modes, and the G-band is nearly as much.

In the next section, we discuss the method, and in the following sections we discuss our results, ending with the conclusions that that the RBM is the most anharmonic of the vibrational modes of the SWNT as revealed by its shift with temperature. We also find that the anharmonicity of these modes is not sensitive to chirality or length but does depend on diameter. We discuss our results in relation to experiments and previous calculations.

\section{Method}

The moments method is an approximation based on low-order moments of
the Liouvillian operator~\cite{momentstheory}, which is the
time-evolution operator of phase-space functions for a \emph{classical} dynamical system. Beginning
with the harmonic force constant matrix for the particular cell, the normal modes are found,
indexed by wavevector $k$ and branch $b$. The calculation involves ensemble averaging of products of normal mode amplitudes $A_{kb}$ and accelerations $\ddot{A}_{kb}$, which are obtained by projecting the atomic displacements and forces onto the normal modes. Using the harmonic modes as a basis is justified by the weakly anharmonic character of this system. The lowest, non-trivial moment of the power spectrum of the displacement-displacement autocorrelation 
\begin{equation}
\mu_2(kb) = -\frac{\langle A_{kb} \ddot{A}_{kb} \rangle}{\langle A_{kb}^2 \rangle - \langle A_{kb} \rangle ^ 2}
\label{eq:mu2}
\end{equation}
(where the angle brackets indicate ensemble averages) gives a simple measure of the temperature-dependent dynamics of the system.

The previous expression includes the possibility that $\langle A_{kb} \rangle$ is non-zero, which was not included in our previous papers because the systems considered previously had sufficiently high symmetry that the average displacement vanished. However, in the present case, the average displacement of the RBM deviates significantly from zero as the temperature increases, and so we have extended the expressions derived in previous work to include non-zero first moments.

The quasi-harmonic (temperature-dependent) frequency $\omega(k)$ is given by 
\begin{equation}
\omega(kb) = \sqrt{\mu_2(kb)} 
\label{eq:omega}
\end{equation}
The moment is calculated by standard Monte Carlo integration. This method was used recently to study the anharmonic renormalization of flexural modes in graphene.~\cite{Wang2016} 

In this study, the interatomic interaction is described by a Tersoff potential tailored somewhat for graphene~\cite{TersoffPRL,origTersoff,modTersoff}. The normal modes are identified from the eigenvectors of the harmonic force constant matrix. We generated various tubes of different chirality, diameter and length. (The diameter is related to the chirality by $d = \sqrt{n_1^2+n_1 n_2+n_2^2}\ a_0/\pi$ in which ($n_1$,$n_2$) are the usual chiral indices and $a_0$ = 2.46 \AA, the in-plane lattice constant of graphene.~\cite{Kurti}) We studied tubes with lengths ranging from 1.3 $nm$ to 13 $nm$ and diameters from 0.78 $nm$ to 2.35 $nm$, with various chiralities. All computational cells are under periodic boundary condition along the length of the tube. The Monte Carlo calculation includes 40,000 steps per atom. The lattice parameter of the cells is determined at each temperature by molecular dynamics with adjustable cell size at constant (zero) pressure, thus incorporating thermal expansion or contraction. In some parts of the analysis, we focus on particular modes, but all have been included in the calculation.

As noted above, the RBM is highly symmetrical; the displacements are dominantly radial and are uniform along the length of the tube. Other modes displacing along the radial direction --- we refer to simply and more generally as ``radial modes'' --- are related to the RBM by a dispersion along the tube axis --- as in $\cos{(k z)}$ where $k$ is the wavevector and $z$ is the distance along the axis of the tube ---  and/or a phase related to the azimuth angle $\phi$  --- as in $\cos{(m \phi)}$ with integer $m$. The $k=0$ modes are at the $\Gamma$-point in the one-dimensional Brillouin zone. The RBM is then a special radial mode with $m=0$ and $k=0$. Most of the radial modes are not visible in Raman spectra, but as we have noted before the RBM appears prominently. 

\section{Results}

\begin{figure} \includegraphics[width=\columnwidth]{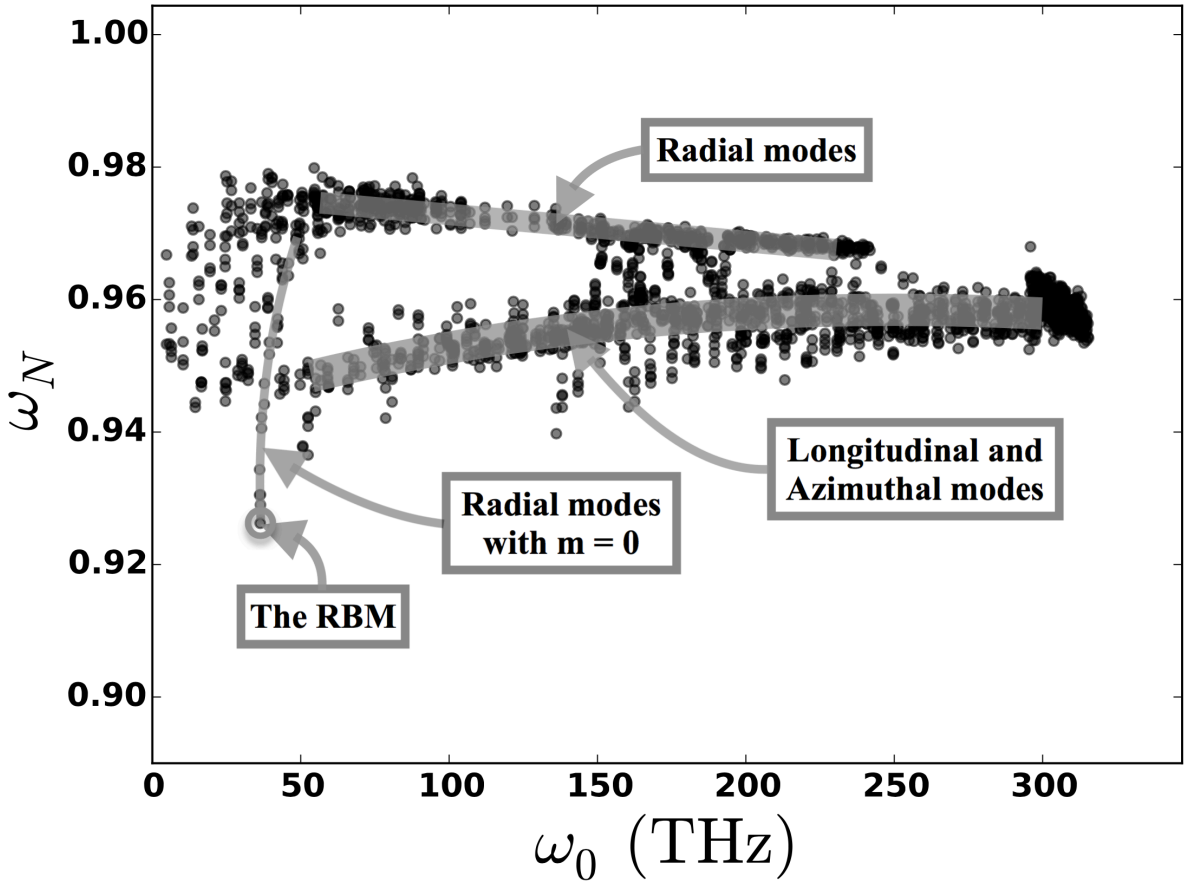} \caption{\label{fig:scatter} Scatterplot of the ratio $\omega_N \equiv \omega/\omega_0$ vs. $\omega_0$ for all of the modes of a SWNT(15,0) tube. $\omega$ is the frequency at temperature ($T=1200K$) and $\omega_0$ is the corresponding harmonic frequency (at $T=0K$). For a perfectly harmonic system, all points would be at the top (at a value of 1); due to anharmonicity, all of the modes of this system drop to lower frequency with increasing temperature. The more anharmonic modes exhibit lower values of $\omega_N$. Lines are drawn in this figure as guides to the eye to indicate the different groupings of modes described in the text. }
\end{figure}

We begin by presenting the results for a typical nanotube. The anharmonicity is easily displayed by comparing the frequency $\omega$ at some temperature as compared to the zero-temperature frequency ($\omega_0$). Figure~\ref{fig:scatter} is a scatterplot of the ratio $\omega_N \equiv \omega/\omega_0$ vs. $\omega_0$ for all modes of a SWNT(15,0) nanotube with periodic length 6.5 nm (so it has 900 atoms in the unit cell) at $T=1200K$.  As can be seen from the figure, closer analysis shows that the modes fall into three basic groups according to anharmonicity. The least anharmonic modes (those that show the least drop in frequency with increasing temperature) are the radial modes (except for a small group which will be discussed in a moment). The modes with largely longitudinal and azimuthal displacements are more anharmonic. The modes with displacements mostly in-plane would seem to be more anharmonic because displacements in the plane will have a more direct effect on the bond length than out-of-plane displacements.

\begin{figure} \includegraphics[width=\columnwidth]{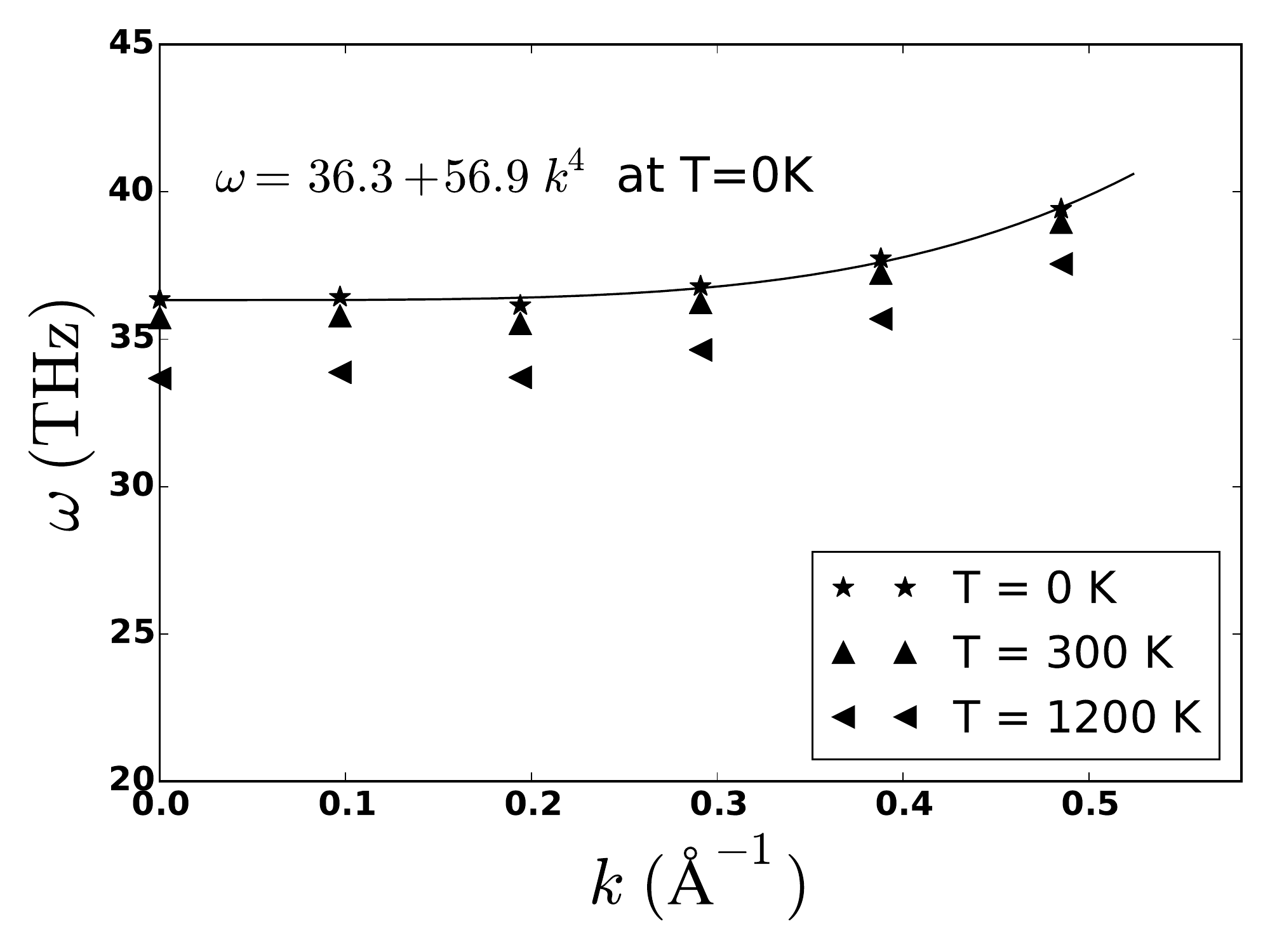} \caption{\label{fig:tdispersions} The dispersion relations of the azimuthally symmetric (that is, $m=0$), radial modes of SWNT(15,0) at various temperatures calculated using the moments method. The periodic length is 6.5 nm, so there are 900 atoms in the unit cell.  A fit to Eq.~\ref{eq:quartic} is shown. The results for several other intermediate temperatures have been calculated and are intermediate to these results, so they have been omitted from this plot for the sake of presentation. }
\end{figure}

\begin{figure}
\includegraphics[width=\columnwidth]{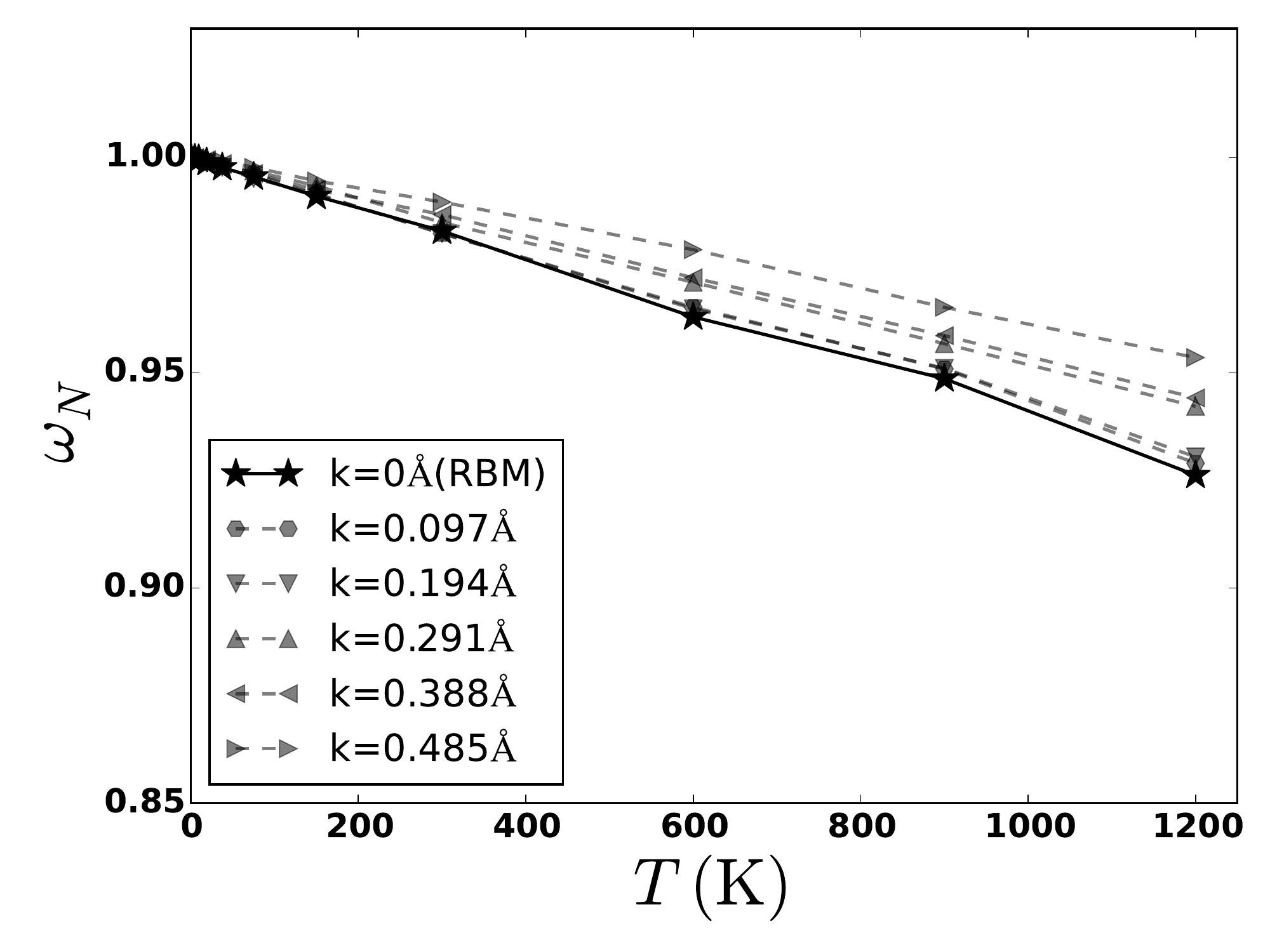}
\caption{\label{fig:ratiovsT} The ratio of anharmonic frequency to harmonic frequency ($\omega_N(T) = \omega(T)/\omega_0$) vs temperature for radial modes of various wavevector $k$. }
\end{figure}

Finally, in Fig.~\ref{fig:scatter} there is a stand-out set at a frequency of about 35 THz, which are radial modes that are azimuthally symmetric (that is, have $m=0$). These modes include the RBM and related modes with $k \neq 0$. Among this set, we find that the RBM is the most anharmonic of all of the modes of the nanotube. This is consistent with the experimental Raman results reported in Ref.~\cite{Raravikar2002}, where both the RBM feature and the G-band are seen to shift with temperature but in fractional terms the RBM is more anharmonic. These azimuthally symmetric radial modes differ from the RBM by the longitudinal wavenumber $k$. So we focus on these modes by showing the temperature-dependent dispersion curves in Fig.~\ref{fig:tdispersions}.  
At $T=0K$ the frequencies (purely harmonic) follow a quartic dispersion
\begin{equation} \label{eq:quartic} \omega(k) = c_0 + c_1 k^4 \end{equation}
The value of $c_0$ (at $k=0$) then corresponds to the frequency of the RBM. The temperature dependence is made clearer in Fig.~\ref{fig:ratiovsT} by replotting the same data in terms of the ratio to the harmonic frequency vs. temperature. 
In both figures, it is clear that the RBM has stronger temperature dependence than the other azimuthally symmetric radial modes.  The frequency of the RBM is roughly linear with temperature, and the temperature coefficient $p$ in
\begin{equation}
\label{eq:p}
\omega_N(T) = 1  + p T 
\end{equation}
is determined for this tube to be $-6 \times 10^{-5} K^{-1}$. Fig.~\ref{fig:ratiovsT} also shows that $p$ decreases with increasing wavevector $k$. The RBM is the most anharmonic of the set (indeed, of all modes).

Two different experimental observations using Raman spectra of nanotube powders~\cite{HDLi2000, Raravikar2002} of similar diameter (both estimated to be 1.3 to 1.4 nm) determine a slope in the range of 300-800 $K$ of $-7 \times 10^{-5} K^{-1}$ and $-2 \times 10^{-5} K^{-1}$, which span our value. We note that it is not clear how to compare calculations of isolated nanotubes with the experimental Raman spectra of nanotube powders; it is expected that in such powders the tubes touch and interactions with neighboring tubes would affect their vibrational characteristics. This could be studied theoretically by bringing tubes in contact with each other, which is beyond the scope of the present calculation, but could be done in the future with the same methods.

A Molecular Dynamics (MD) simulation~\cite{Raravikar2002} using one of the original Tersoff potentials for carbon~\cite{TersoffPRL} for a (10,10) tube (1.4 nm) reported a value of $p$ of $-5 \times 10^{-5} K^{-1}$, very close to our calculation and also within the range reported in the two experiments. The minor difference between the present calculations and those previously reported might be attributed in part to the different chiralities and diameters considered, in part to the interatomic potential (we used a potential with Tersoff form that was somewhat optimized to graphene~\cite{modTersoff}), or possibly in part to the difference in technique (MD vs. moments). At any rate, the differences are not large.

One might hypothesize that the RBM is more anharmonic than other modes because it is radial --- that is, maybe displacements in the radial direction are more anharmonic than displacements in other directions --- but that is disproved by our results that demonstrate that the large majority of radial modes are not as anharmonic as the RBM. The strong temperature dependence of the RBM frequency, especially in contrast to the other radial modes with $k \neq 0$ can be understood rather in terms of the high symmetry of the RBM (with $k=0$ and $m=0$). The RBM has the same symmetry as the nanotube itself  (in that any operation that returns the atomic structure of the nanotube to itself also returns the RBM to itself) and can therefore couple easily with other modes. In terms of the usual 3rd order and 4th order force constant matrices (see, for example, Madelung~\cite{Madelung}), the high symmetry of the RBM allows it to couple with many other modes of the tube. This ubiquitous coupling would mean that the frequency shift of the RBM may depend on the number of other modes present in the tube. The number of such modes will vary with tube diameter and length, so we investigate that here. 

\begin{figure}
\includegraphics[width=\columnwidth]{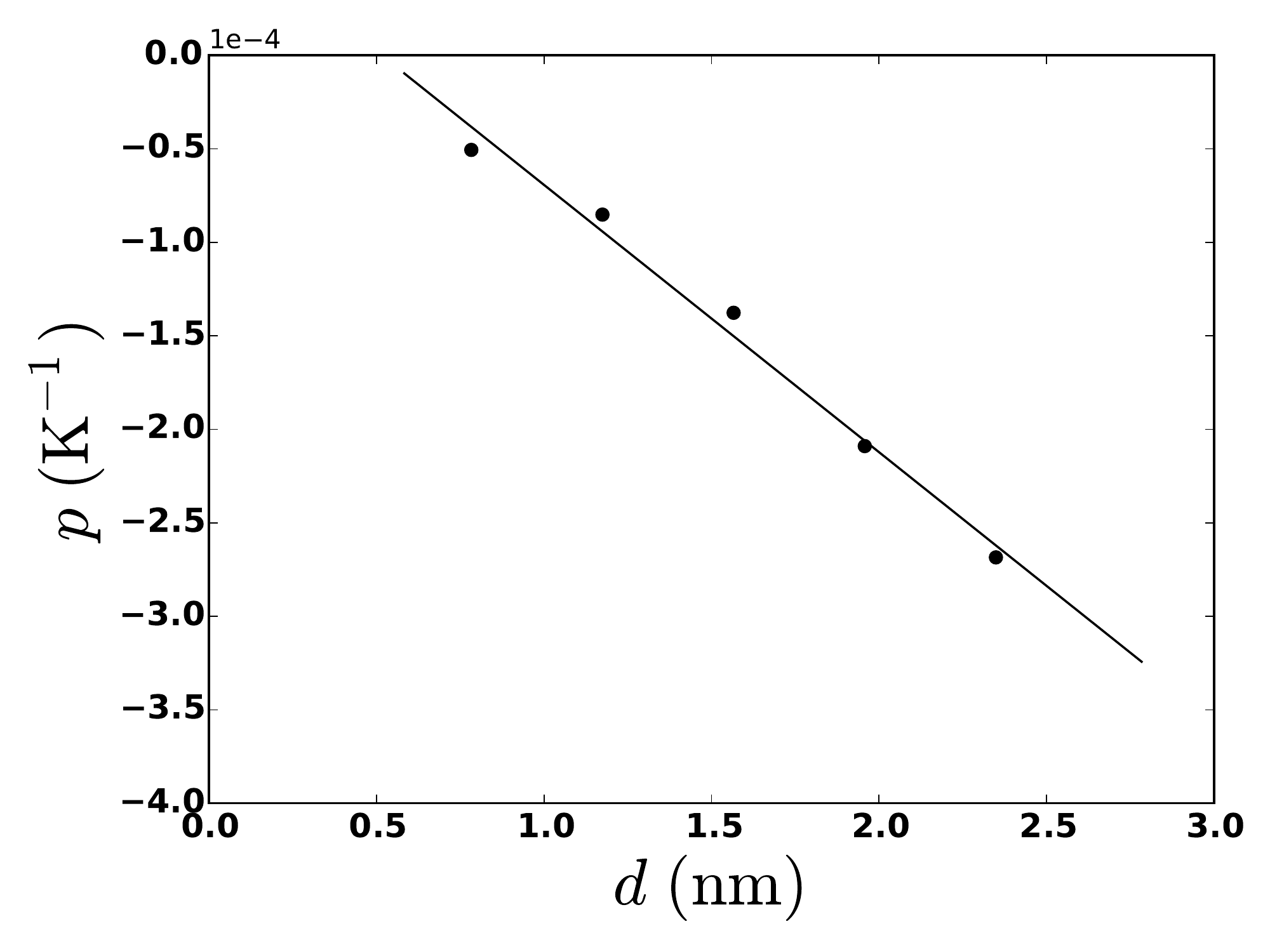}
\caption{\label{fig:slopevsdiameter} The slope $p$ of the temperature dependence of the frequency (Eq.~\ref{eq:p}) of the RBM is plotted vs. the diameter $d$ of the tube. For this plot, we chose a series of tubes with chirality from (10,0) to (30,0). Other corresponding sequences of tubes show similar behavior.}
\end{figure}

In Fig.~\ref{fig:slopevsdiameter},
we show that the value of $p$ for the RBM (from Eq.~\ref{eq:p}, Fig. ~\ref{fig:ratiovsT}) depends on the diameter of the tube. That is, the anharmonicity of the RBM is stronger for tubes with larger diameter. This is consistent with the observation in the previous paragraph, because as the diameter of the tube increases the number of modes that are available to couple with the RBM increases, thereby resulting in a stronger anharmonicity. 

\begin{figure}
\includegraphics[width=\columnwidth]{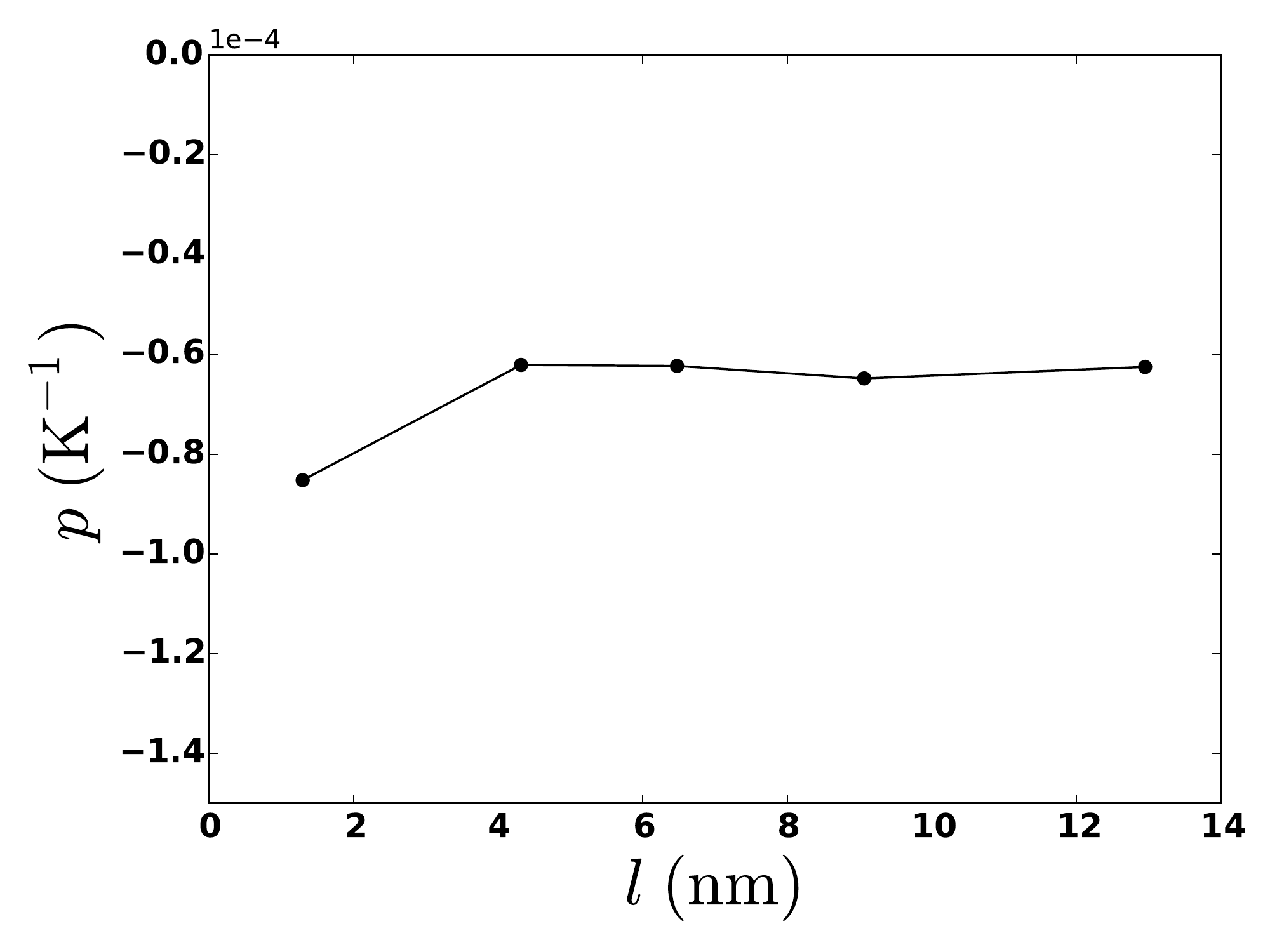}
\caption{\label{fig:slopevslength} The slope $p$ of the temperature dependence of the frequency (Eq.~\ref{eq:p}) of the RBM of the (15,0) SWNT is plotted vs. the periodic length $l$ of the supercell. }
\end{figure}

However, by contrast, we show in Fig.~\ref{fig:slopevslength} that the temperature dependence of the modes is not very dependent on the periodic length of our calculation. Periodicity restricts the calculation so that only vibrational modes with wavelength commensurate with the periodic length are allowed. Increasing the length of the periodic cell allows more modes to be present, and also to couple with the RBM which is present always. However, our results indicate only a weak dependence of the anharmonicity on the periodic length.

\section{Conclusions}

We have presented the results of a study of anharmonicity of vibrational modes of SWNTs obtained by the ``moments method'', which is based on Monte Carlo averages of products among displacements and forces. The forces and energies required for the MC calculation were obtained from a semi-empirical Tersoff potential, somewhat optimized for graphene. 

Generally all modes shift down in frequency with increasing temperature. Modes with largely in-plane character (longitudinal and azimuthal modes) are more strongly anharmonic than most modes with radial character, with the exception of azimuthally symmetric radial modes (which includes the RBM). In terms of the fractional shift in frequency, the RBM is the most anharmonic of all modes of the nanotube. The temperature dependence of the frequency of the RBM increases with the diameter of the tube. This is attributed to an increase in the number of azimuthal modes eligible to couple with the RBM. As far as we can determine from searching the literature, this is the first time that the effect of diameter on the anharmonicity has been appreciated. The pronounced anharmonicity of the RBM is attributed to its ubiquitous coupling to other azimuthal modes, which is allowed by its high symmetry. The results provide a clearer picture of the anharmonicity of the vibrational modes of a SWNT.
  
\section{Acknowledgement}
Our research was supported by the US Department of Energy, Office of Basic Energy Science, Division of Materials Sciences and Engineering under Grant No. DE-SC0008487.

\end{document}